\documentclass[pdflatex,sn-mathphys-num]{sn-jnl}

\usepackage{graphicx}%
\usepackage{multirow}%
\usepackage{amsmath,amssymb,amsfonts}%
\usepackage{amsthm}%
\usepackage{mathrsfs}%
\usepackage[title]{appendix}%
\usepackage{xcolor}%
\usepackage{textcomp}%
\usepackage{manyfoot}%
\usepackage{makecell}
\usepackage{array}%
\usepackage{algpseudocode}%
\usepackage{listings}%
\usepackage{makecell}
\usepackage{placeins}
\usepackage{float}
\usepackage{booktabs}
\usepackage{diagbox}
\usepackage{tabularx}
\usepackage{array}
\usepackage{longtable}
\usepackage{caption}
\usepackage{subcaption}
\usepackage{silence}
\usepackage{url}
\usepackage{threeparttable}
\usepackage{diagbox}
\usepackage[linesnumbered,ruled,vlined]{algorithm2e}
\usepackage{epstopdf}

\theoremstyle{thmstyleone}%
\theoremstyle{thmstyletwo}%

\theoremstyle{thmstylethree}%

\begin{document}

\title[Article Title]{Combined machine learning for stock selection strategy based on dynamic weighting methods}

\author[1]{\fnm{Lin} \sur{Cai}}\email{lc3881@columbia.edu}
\author[2]{\fnm{Zhiyang} \sur{He}}\email{zh301@sussex.ac.uk}
\author*[3]{\fnm{Caiya} \sur{Zhang}}\email{zhangcy@hzcu.edu.cn}

\affil[1]{\orgdiv{Department of Statistics}, \orgname{Columbia University}, \orgaddress{\city{New York}, \state{NY}, \postcode{10027}, \unskip~\country{USA}}}

\affil[2]{\orgdiv{Department of Engineering and Informatics}, \orgname{University of Sussex}, \orgaddress{\street{Falmer}, \city{Brighton}, \postcode{BN1 9RH}, \country{United Kingdom}}}

\affil*[3]{\orgdiv{Department of Statistics and Data Science}, \orgname{Hangzhou City University}, \orgaddress{\street{Huzhou Street}, \city{Hangzhou}, \postcode{310015}, \state{Zhejiang}, \country{China}}}


\abstract{This paper proposes a novel stock selection strategy framework based on combined machine learning algorithms. Two types of weighting methods for three representative machine learning algorithms are developed to predict the returns of the stock selection strategy. One is static weighting based on model evaluation metrics, the other is dynamic weighting based on Information Coefficients (IC). Using CSI 300 index data, we empirically evaluate the strategy’s backtested performance and model predictive accuracy. The main results are as follows: 
(1) The strategy by combined machine learning algorithms significantly outperforms single-model approaches in backtested returns.
(2) IC-based weighting (particularly $IC_{Mean}$) demonstrates greater competitiveness than evaluation-metric-based weighting in both backtested returns and predictive performance.
(3) Factor screening substantially enhances the performance of combined machine learning strategies. }



\maketitle

\section{Introduction}\label{sec1}

It is well known that compared to traditional investment based on fundamental analysis or technical indicators, quantitative investment  offers some important advantages such as stable returns, strong verifiability, and the ability to avoid irrational behavior. In quantitative investment, stock selection  based on prediction models plays a crucial role. It can not only diversify the risks in the investment portfolio, but also enhance the adaptability of quantitative investment strategies.

With the rapid development of machine learning, an increasing number of scholars and investors have attempted to apply these prediction techniques to the field of quantitative investment. Lee (2009) applied Support Vector Machines (SVM) to stock classification and portfolio selection, demonstrating that SVM outperformed the Backpropagation Neural Network (BPNN) in predicting stock trends \cite{lee2009using}. Recently, Wang (2022) developed a stock selection model based on 26 factors, using a gated recurring unit (GRU) neural network optimized by the Cuckoo Search (CS) algorithm. This model demonstrated superior backtest performance when compared to other benchmark models\cite{wang2022intelligent}. Wang  proposed a stability prediction model based on contradictory factors, using the sigmoid deep learning paradigm combined with multifactor models. In this model, the sigmoid learning function determines the stability effects of different factors on profitable stock trading, each factor mapped to the profit results of real-time stocks and their trading values\cite{wang2025multifactor}. The results showed that the sigmoid layer, after training, can recognize sudden changes in the stock market.

Note that the aforementioned studies primarily employed single machine learning algorithms, which are limited in their ability to capture diverse data characteristics across different time spans. To address these limitations and enhance prediction performance, the combination of multiple machine learning algorithms has become a widely adopted technique. Nti et al (2020) investigated the application of various ensemble learning methods in stock market prediction\cite{nti2020comprehensive}. In their paper,  Decision Trees (DT), Support Vector Machines (SVM), and Neural Networks (NN) were utilized, and ensemble predictions were conducted using basic techniques such as weighted averaging, majority voting, and averaging, as well as more advanced techniques including stacking, blending, bagging, and boosting. The weighted averaging method was used to combine prediction results by assigning different weights to each model. Mozaffari and Zhang (2024) explored the enhancement of stock price prediction accuracy through ensemble learning methods was explored\cite{mozaffari2024predicting}. Linear Regression, ARIMA, and Transformer models were utilized, and predictions were combined using stacking regressors and weighted ensemble methods. In the same year, Zhu and Wang (2024) examined the classification of ST and non-ST stocks in the Chinese stock market using multiple machine learning models\cite{zhu2024stock}. Logistic Regression, XGBoost, LightGBM, AdaBoost, CatBoost, and Multilayer Perceptron (MLP) models were employed, with XGBoost selected as the optimal model. The AdaBoost algorithm was applied for weighting, where data weights and weak classifier weights were adjusted to improve the accuracy and robustness of the classifier. Across all these studies, regression or classification evaluation metrics were used as the basis for weight adjustment. These methods significantly improved prediction accuracy and robustness, yet certain drawbacks remained. Firstly, high computational complexity was introduced, as multiple model predictions and their weighted combinations had to be calculated. Secondly, there was a risk of overfitting, particularly when handling noisy data. Additionally, changes in market conditions could reduce model adaptability, requiring continuous updates and adjustments to maintain effectiveness under varying market environments.

In our work, three representative types of machine learning algorithms are used, including Ridge Regression, Multilayer Perceptron (MLP) neural networks, and Random Forest. The main challenge lies in determining the weights of prediction outcomes by different algorithms.  At first, we utilize regression evaluation metrics, namely Root Mean Squared Error (RMSE) and Mean Absolute Percentage Error (MAPE), to determine the weights. The target variable of interest is the return. The algorithm with a smaller prediction error will receive a higher weight. The empirical results indicate that it can indeed help reduce the absolute prediction error. However, it cannot predict the direction of price change.
 To overcome this limitation, we resort to the classification evaluation metrics, including precision, recall, and the F1-score. In other words, we take whether the stock price will rise or fall as our prediction target. The algorithm with a higher prediction accuracy ( recall or  F1-score) will be assigned a higher weight.

To combine the advantages of regression evaluation metrics with classification evaluation metrics, we propose a more competitive weighting method based on the information coefficient (IC). Zheng (2021) proposed a multi-factor stock selection and timing model based on IC, and empirical results showed that this method has strong anti-risk capabilities in bear markets, achieving stable excess returns\cite{zheng2021multi}. However, it cannot dynamically adjust the weights according to the  market conditions, thus reducing the model's adaptability. Li (2020) used the XGBoost model to predict the IC value for the next 10 trading days and dynamically adjusted factor weights based on the predictions\cite{li2020overview}. Empirical results showed that its strategy significantly outperformed strategies using equal weights and historical static IC-based weighting, showing better market adaptability. However, the IC values by this method rely on prediction models, the stability of which is affected by market conditions, data quality, and parameter choices.

To address these limitations, this paper proposes two real-time dynamic weighting methods. Firstly, we calculate the IC values through Spearman correlation between predicted returns and real returns, which can reflect both the direction and magnitude of the changes in stock price. Then two IC-based weighting schemes are provided. The first is to calculate the average IC value ($IC_{Mean}$) over a fixed trading time rolling window . The second is to adjust the $IC_{Mean}$ by its
standard deviation, denoted by $IC_{Ratio}$, to account for the stability of predictive performance. Unlike Zheng's (2021) static weighting approach, our method dynamically adjusts factor weights using rolling historical data, ensuring adaptive optimization in response to market fluctuations\cite{zheng2021multi}. Additionally, in contrast to Li's (2020) method, our approach calculates IC values independently of machine learning models, thereby enhancing the robustness and interpretability of the weights\cite{li2020overview}. Empirical results demonstrate that the proposed IC-based dynamic weighting strategy outperforms competing methods in terms of strategy returns, model robustness, and market adaptability. Specifically, the strategy's returns are nearly four times higher than those of a single machine learning algorithm and more than double those of evaluation metric-based weighting.

It is also worth mentioning the issue of feature selection. This study draws inspiration from multi-factor models. The multi-factor model,as the earliest and most widely used quantitative stock selection model, is based on the core idea of identifying the most relevant factors that influence stock returns and selecting stocks accordingly. In 1974, Rosenberg introduced an important type of multi-factor model for the US equity market, Barra model\cite{rosenberg1974extra}. This model provided investors with a scientific and systematic approach to quantitative stock selection. It incorporates various factors, such as country factors, industry factors, and style factors. Investors can choose appropriate factors to construct a stock-picking model based on the characteristics of the target market. With the widespread promotion and application of the Barra model globally, Morgan Stanley Capital International (MSCI) introduced the Barra China Stock Model, specifically known as the CNE model, which is tailored to the Chinese equity market. In June 2012, MSCI released the CNE5 version, marking a significant milestone in the development of the CNE model\cite{MSCI2012CNE5}. In the CNE5 model, besides the country factor and industry factors, there are 10 types of style factors including  capitalization factor, beta factor, momentum factor, and profitability factor so on. At present, the CNE5 model has become an indispensable tool for risk management and portfolio optimization in China's capital market. 

With the development of financial markets and the advancement of technology, new concepts and investment strategies have continuously emerged. The CNE5 factor model has been criticized for its structural rigidity, limited adaptability to evolving market conditions, and over-reliance on traditional financial indicators. These shortcomings have constrained its effectiveness in capturing emerging risk premia and investor behavior. To improve return stability, enhance the probability of outperforming industry averages, and capture the reversibility of stock returns, we construct 17 new second-level style factors adding to the CNE5 model. The empirical results of factor screening based on LASSO with the data from the CSI 300 index  validate the importance of these new factors in return forecasting.

The rest of the paper is structured as follows. New Barra factors are constructed and factor screening is conducted in Section \ref{sec2}. In Section \ref{sec3}, stock selection strategies based on three single machine learning algorithms and seven combined predictors are discussed in detail. Finally, conclusion and discussion are presented in Section \ref{sec4}.

\section{Factor Construction and Screening}
\label{sec2}

\subsection{Data Preprocessing}

In this study, the constituent stocks of the CSI 300 Index are selected as the stock pool, with monthly sample data spanning from January 1, 2018, to December 31, 2022\cite{tushare2023data}. To ensure data validity, stocks labeled as ST (Special Treatment), suspended, or newly listed within the past month are excluded.

For missing values in certain factors, imputation is performed using the industry median based on the Shenwan Industry Classification Standard\cite{swsresearch2023data}. When the industry median is unavailable, the overall median of the stock pool is applied. Outliers are detected using the three times Median Absolute Deviation (MAD) rule; observations exceeding three times the MAD from the sample median are treated as outliers. All factor values are then standardized using Z-score normalization.

Furthermore, to eliminate the influence of industry and market capitalization effects on stock performance, these variables are removed from the factor set to better capture firm-specific characteristics. A regression-based residual centering method is adopted. Specifically, each style factor is regressed on industry dummies and market capitalization, and the residuals from the regression are retained as the adjusted factor values. For factors inherently related to market capitalization (e.g., the market capitalization factor itself), only industry centering is performed.

\subsection{Factor Construction}

Drawing on asset pricing theory (APT), behavioral finance, and corporate finance theory, our study initially selects eight style factors from the Barra China Stock Model (CNE5): \textit{Size}, \textit{Reverse}, \textit{Volatility}, \textit{Earning}, \textit{Growth}, \textit{Value}, \textit{Leverage}, and \textit{Liquidity}. These primary factors are then expanded into a total of 33 second-level factors. Furthermore, we develop 17 new second-level factors in this paper with the aim of augmenting the portfolio's potential for generating excess returns.

\textit{(1) Size}: This factor primarily gauges the magnitude of a stock's market capitalization and serves as a fundamental component in the Fama-French Three-Factor Model\cite{fama1993common}. According to this model, small-cap stocks often exhibit a tendency to generate excess returns over extended periods. It encapsulates the market's inclination towards size premiums, frequently linked to liquidity risks and information asymmetry. The newly introduced factor, \textit{IR\_industry\_total\_mv}, quantifies the consistency of a stock's total market value in relation to the industry average. The information ratio (IR), which serves as an indicator of the factor's stability, is derived by dividing the average value by which a stock's factor surpasses the industry mean over the preceding 20 trading days by its standard deviation.

\textit{(2) Reverse}: This factor plays a pivotal role in elucidating variations in asset returns, as it delineates the reversal of short-term price trajectories. The Overreaction Hypothesis within behavioral finance posits that investor sentiment and market perturbations can precipitate transient deviations from intrinsic values. Notably, Jegadeesh and Titman (1993) discovered that reversal strategies frequently generate superior alpha returns during periods of market turbulence or crises\cite{Narasimhan1993returns}. To comprehensively capture reversal effects across a spectrum of dimensions—encompassing returns, turnover rates, volatility, trading volumes, and capital flows—we introduce eight novel reversal-related factors: \textit{Pct\_chg} (percentage change in returns), \textit{Turnover\_rate} (rate of stock turnover), \textit{Amplitude} (magnitude of price fluctuations), \textit{Vol} (volatility), \textit{Sm} (short-term reversal factor), \textit{Md} (medium-term reversal factor), \textit{Lg} (long-term reversal factor), and \textit{Elg} (extended long-term reversal factor).

\textit{(3) Volatility}: This metric captures the fluctuations in asset prices and acts as a surrogate indicator for risk levels. In alignment with Modern Portfolio Theory (Markowitz, 1952), volatility stands as a pivotal yardstick for gauging investment risk\cite{Markowitz1952Portfolio}. Elevated volatility typically signals heightened market uncertainty and the presence of risk premiums. Additionally, the low-volatility anomaly underscores that stocks with lower volatility often yield superior risk-adjusted returns, thereby posing a challenge to conventional financial theories.

\textit{(4) Earning}: As a pivotal factor in elucidating asset returns, it underscores the pivotal role that corporate profitability plays in shaping stock performance. Graham and Dodd (1934) underscore its significance within the realm of value investing. Profitability serves as a bedrock indicator of a firm's ability to generate value and exerts a direct influence on its intrinsic valuation\cite{Benjamin1934Security}. The Fama-French Five-Factor Model (Fama and French, 2015) further cements profitability's status as a core pricing factor within its framework\cite{fama2015five}.

\textit{(5) Growth}: This factor quantifies a company's growth potential based on market expectations regarding its future performance. According to shareholder value theory, the present stock price is fundamentally determined by the anticipated future cash flows. Growth stocks are predominantly found in high-growth sectors, including technology and consumer goods. Asness, Frazzini, and Pedersen (2019) have demonstrated that the growth factor acts as a complement to the value factor, particularly pronounced during periods of economic expansion\cite{asness2019quality}. To evaluate the stability of profitability, we have devised two novel factors—\textit{IR\_eps} and \textit{IR\_revenue\_ps}—which are computed as the ratio of the mean to the standard deviation over a five-year timeframe.

\textit{(6) Value}: This factor gauges the relationship between asset prices and valuation metrics. Rooted in value investing theory, it posits that undervalued stocks are poised to deliver superior long-term performance. The HML (High Minus Low) factor within the Fama-French model (Fama and French, 1993) provides empirical validation for the valuation effect\cite{fama1993common}. To evaluate the stability of valuation in comparison to industry counterparts, we introduce two novel factors: \textit{IR\_industry\_pb} and \textit{IR\_industry\_ps}. These factors encapsulate the information ratio for the price-to-book (P/B) and price-to-sales (P/S) ratios, respectively, thereby offering insights into the relative valuation stability within the industry.

\textit{(7) Leverage}: This factor assesses the impact of financial leverage on stock returns. Although the Modigliani-Miller theorem (Modigliani and Miller, 1958) posits that, under ideal conditions, a firm's capital structure should have no bearing on its value, real-world imperfections result in leverage magnifying both returns and risk\cite{modigliani1958cost}. Baker and Wurgler (2019) have noted that leverage ratios exhibit an inverse relationship with beta, a phenomenon that cannot be entirely accounted for by conventional theories of financial distress\cite{baker2020leverage}. To gauge the temporal stability of a firm's leverage, we propose three new indicators: \textit{IR\_mlev}, \textit{IR\_dtoa}, and \textit{IR\_blev}.

\textit{(8) Liquidity}:  This factor captures the liquidity characteristics of a stock, reflecting the level of investor demand and the ease with which it can be traded. Amihud and Mendelson (1986) pioneered the liquidity discount theory, which posits that illiquid assets necessitate higher expected returns as compensation for the associated liquidity risk \cite{amihud1986asset}. Pastor and Stambaugh (2019) have further substantiated that liquidity risk premiums tend to escalate markedly during periods of market downturn \cite{pastor2019liquidity}. To quantify the consistency of liquidity at the industry level, we have developed a novel factor—\textit{IR\_industry\_turnover\_rate}—which represents the information ratio of a stock's turnover rate in relation to the industry average.

 The definitions of 50 second-level style factors are presented in Table \ref{factor_definition}.

\subsection{Factor Selection and Synthesis}
 The presence of multicollinearity may lead to unstable parameter estimates, diminished explanatory power, and a heightened risk of overfitting as well. To mitigate this issue, it is necessary to conduct a factor selection. The Least Absolute Shrinkage and Selection Operator (LASSO), proposed by Tibshirani (1996), is employed here \cite{tibshirani1996regression}. By incorporating an $L_1$ regularization term into the regression objective, LASSO enables automatic factor selection and coefficient shrinkage, which contributes to improved model robustness and predictive accuracy. The optimization objective function is formulated as follows.
\begin{equation}
\min_{\beta} \left( \sum_{i=1}^{n} (y_i - \sum_{j=1}^{p} X_{ij} \beta_j)^2 + \lambda \sum_{j=1}^{p} |\beta_j| \right)
\end{equation}
where $y_i$ is the stock return, $X_{ij}$ is the $j$-th factor, and $\lambda$ is the regularization parameter. LASSO adjusts $\lambda$ to balance the number of factors and the model’s fit, retaining only the factors that contribute significantly to return prediction.

Through LASSO method , a total of 13 second-level style factors  are excluded, only including 2 newly constructed factors, which are the reversal factor \textit{Vol} and the growth factor \textit{IR\_revenue\_ps}. The detailed results of factor selection can be seen in Table \ref{factor_definition}.

To quantify the importance of each second-level indicator under a given first-level factor, we adopt the Entropy Weight Method (EWM) that reflects the dispersion of indicator values across a set of stocks. The basic idea of EWM is to assign weights according to the entropy values of each indicator. The higher the entropy value, the greater the weight it should have, and vice versa. Let \( x_{ijt}^{a} \) be the value of the \( j \)th second-level indicator among the \( i \)th first-level factor, for stock \( a \) at time \( t \). The weighting process follows four key steps.

\textbf{Step 1: Standardization} \\
Each indicator value is scaled to the \([0, 1]\) range using min-max normalization over the past 12 months.
\begin{equation}
z_{ijt}^{(a)} = 
\frac{
	x_{ijt}^{(a)} - \min\limits_{\substack{s \in [t-12,\, t-1] \\ b\in [1, N]}} x_{ijs}^{(b)}
}{
	\max\limits_{\substack{s \in [t-12,\, t-1] \\ b \in [1, N]}} x_{ijt}^{(b)} -
	\min\limits_{\substack{s \in [t-12,\, t-1] \\ b \in [1, N]}} x_{ijt}^{(b)}
}
\tag{1}
\end{equation}

\textbf{Step 2: Normalization} \\
\begin{equation}
p_{ijt}^{(a)} = \frac{z_{ijt}^{(a)}}{\sum_{s=t- 12}^{t-1} \sum_{a = 1}^{N} z_{ijt}^{(a)}}
\tag{2}
\end{equation}

\textbf{Step 3: Entropy Calculation} \\
The entropy for the \( j \)th indicator under the \( i \)th first-level factor at time \(t\) is as follows.
\begin{equation}
e_{ijt} = -\frac{1}{\ln(12 \cdot N)} \sum_{s = t-12}^{t-1} \sum_{a = 1}^{N} p_{ijs}^{(a)} \cdot \ln p_{ijs}^{(a)}
\tag{3}
\end{equation}

\textbf{Step 4: Weight Determination} \\
The weight for the \( j \)th indicator under the \( i \)th first-level factor at time \(t\) is determined as
\begin{equation}
w_{ijt}= \frac{1 - e_{ijt}}{\sum\limits_{j=1}^{m_i} \left(1 - e_{ijt}\right)}
\tag{4}
\end{equation}

Then the aggregated score for stock \( a \) with respect to the \( i \)th first-level factor is given by
\begin{equation}
F_{it}^{(a)} = \sum_{j=1}^{m_i} w_{ijt} \cdot z_{ijt}^{(a)}
\tag{5}
\end{equation}

A rolling window approach is employed to dynamically update the factor weights and enhance their adaptability to changing market conditions. Specifically, data from the preceding 12 months is used to compute the initial set of weights, which are then applied to the same period’s data. Beginning from the 13th month, the window advances on a monthly basis, and factor weights are recalculated using the most recent 12-month data. The updated weights are then used to generate the first-level factor values for the current month. Through this process, factor weights adjust dynamically over time, thereby mitigating the degradation in factor effectiveness due to structural shifts in the market. The average weights of the second-level indicators are reported in Table~\ref{factor_definition}.

\setlength{\abovecaptionskip}{0cm}
\renewcommand\arraystretch{1.5}
\scriptsize
\begin{longtable}{p{0.15\textwidth}p{0.25\textwidth}p{0.55\textwidth}}
    \caption{Definitions and weights of style factors }\label{factor_definition}\\
    \toprule
    First-level-factor & Second-level-factor (Weight) & Factor Meaning\\
    \midrule
    \endfirsthead

    \multicolumn{3}{l}%
    {{\bfseries \tablename\ \thetable{} -- continued from previous page}} \\
    \toprule
    First-level-factor & Second-level-factor (Weight) & Factor Meaning\\
    \midrule
    \endhead

    \bottomrule
    \endfoot

    \bottomrule
    \endlastfoot
    
    \multirow{4}{*}{\makecell[c]Size} & Total\_mv (0.560261) & Total market value\\
    & Total\_mv\_log (0.000) & Log of total market value\\
    &  Total\_mv\_3 (0.000) & Total market value in cubic meters\\
    & $\triangle$IR\_industry\_total\_mv (0.439739) & Total market value industry excess IR, with a time window of 20 trading days\\

    \midrule
    \multirow{11}{*}{\makecell[c]Reverse} & Roe20 (0.085568) & One-month index-weighted cumulative daily rate of return, with a half-life of 60 trading days\\
    & Roe60 (0.083371) & Three-month index-weighted cumulative daily rate of return, with a half-life of 60 trading days\\
    & Roe120 (0.08265) & Six-month index-weighted cumulative daily rate of return, with a half-life of 60 trading days\\
    & $\triangle$Pct\_chg (0.115577) & If the stock return is less than the market average, multiply the return by -1\\
    & $\triangle$Turnover\_rate (0.112719) & If the stock turnover rate is less than the market average, multiply the return by -1\\
    & $\triangle$Amplitude (0.12456) & If the stock amplitude is less than the market average, multiply the return by -1\\
    & $\triangle$Vol (0.000) & If the stock volume is less than the market average, multiply its return by -1\\
    & $\triangle$Sm (0.100792) & Small order money flow reversal factor \\
    & $\triangle$Md (0.09853) & Middle order money flow reversal factor\\
    & $\triangle$Lg (0.096652) & Large order money flow reversal factor\\
    & $\triangle$Elg (0.09958) & Extremely large order money flow reversal factor\\

    \midrule
    \multirow{3}{*}{\makecell[c]Volatility} & Beta (0.250511) & Perform a time series regression of the stock's index-weighted daily returns against the returns of the CSI 300 Index \\
    & Std\_e (0.344108) & Regression residuals of the stock's index-weighted daily returns \\
    & Std\_g (0.405382) & Volatility of index-weighted daily returns over the past 250 trading days \\
    \multirow{1}{*}{\makecell[c]Volatility}& Dastd (0.000) & Volatility of index-weighted daily excess returns over the past 250 trading days\\

    \midrule
    \multirow{4}{*}{\makecell[c]Earnings} & Etop (0.193039) & Net profit for the last 12 months divided by current total market value\\
    & Rtop (0.000) & Revenue for the last 12 months divided by current total market value\\
    & Epstop (0.352521) & Earnings per share for the last 12 months divided by current total market value\\
    & Cetop (0.45444) & Operating cash flow for the last 12 months divided by current total market value\\
    
    \midrule
    \multirow{9}{*}{\makecell[c]Growth} & Basic\_eps\_yoy (0.000) & Year-over-year growth rate of basic earnings per share\\
    & Netprofit\_yoy (0.169285) & Year-over-year growth rate of net profit \\ 
    & Ocf\_yoy (0.259548) & Year-over-year growth rate of net cash flows from operating activities \\
    & Roe\_yoy (0.167372) & Year-over-year growth rate of return on equity (diluted)\\
    & Or\_yoy (0.191074) & Year-over-year growth rate of operating income\\
    & CGR\_eps (0.000) & CGR of earnings per share over the past five years\\
    & CGR\_revenue\_ps(0.000) & CGR of operating income per share over the past five years \\ 
    & $\triangle$IR\_eps (0.212721) & Earnings per share IR over the past five years\\
    & $\triangle$IR\_revenue\_ps(0.000) & Operating income per share IR over the past five years\\

    \midrule
    \multirow{5}{*}{\makecell[c]Valuation} & Pe (0.000) & Price-earnings ratio (dynamic)\\
    & Pb (0.175991) & Price-to-book ratio\\
    & Ps (0.237718) & Price-to-sales ratio\\
    & $\triangle$IR\_industry\_pb (0.302175) & Pb industry excess IR, with a time window of 20 trading days\\
    & $\triangle$IR\_industry\_ps (0.284117) & Ps industry excess IR, with a time window of 20 trading days\\

    \midrule
    \multirow{6}{*}{\makecell[c]Leverage} & Mlev (0.000) & Market value leverage: (total market value + non-current liabilities)/total market value\\
    & Dtoa (0.129587) & Debt-to-asset ratio: total debt to total assets\\
    & Blev (0.000) & Book leverage: (net assets + non-current liabilities)/net assets\\
    & CGR\_mlev (0.174783) &CGR of market value leverage over the past five years\\
    & CGR\_dtoa (0.200616) & CGR of debt-to-asset ratio over the past five years\\
    & CGR\_blev  (0.000)& CGR of book leverage over the past five years\\
    \multirow{3}{*}{\makecell[c]Leverage} & $\triangle$IR\_mlev (0.156218)& Market value leverage IR over the past five years\\
    & $\triangle$IR\_dtoa (0.13935) & Debt-to-asset ratio IR over the past five years\\
    & $\triangle$IR\_blev (0.199446) & Book leverage IR over the past five years\\
    
    \midrule
    \multirow{4}{*}{\makecell[c]Liquidity} & Turnover\_month (0.207656) & Average monthly turnover rate\\
    & Turnover\_quater (0.226838) & Average quarterly turnover rate\\
    & Turnover\_year (0.267093) & Average annual turnover rate\\
    & $\triangle$IR\_industry\_turnover \_rate(0.298413) & Turnover rate industry excess IR, with a time window of 20 trading days\\

\end{longtable}

\begin{flushleft}
\textbf{Remarks:} The symbol $\triangle$ indicates that the factor is newly constructed and a weight of 0.000 indicates that the factor is excluded.  
\end{flushleft}

\section{Stock Selection  Strategy Based on Machine Learning }
\label{sec3}

\subsection{Forecasting with Single Machine Learning Algorithm}

\normalsize

Three representative machine learning algorithms are selected to construct stock selection strategies: Ridge Regression (Ridge), Multi-Layer Perceptron (MLP) Neural Network and Random Forest. Ridge is a regularized linear regression technique that addresses multicollinearity through the inclusion of an $L_2$ penalty term. MLP is a feedforward artificial neural network composed of multiple layers of interconnected neurons. Random Forest is an ensemble learning method based on decision trees, which aggregates predictions from multiple base learners to enhance model robustness and generalization.

To rigorously assess the predictive performance and generalization ability of stock selection models, a rolling forecast framework based on training and testing is employed. The procedure consists of the following steps.

\textbf{Step 1: Division of training and test set.}
During each prediction cycle, the training set is constructed using historical data from the previous 12-month period. The month that directly follows this 12-month window is then set aside as the test set, specifically for out-of-sample prediction. This approach effectively mimics the decision-making process of investors who base their actions solely on past historical information.

\textbf{Step 2: Model Training and Prediction.}
Each model—Ridge, MLP, and Random Forest—is trained independently using the training dataset. Once trained, the model is utilized to forecast the returns of stock for the upcoming month, relying on the test dataset for these predictions.

\textbf{Step 3: Rolling Window Advancement.}
Following each prediction cycle, the training and test windows are advanced by one month. This rolling mechanism enables continuous updating and adaptation to potential structural changes in the market.

\textbf{Step 4: Return Adjustment and Performance Evaluation.}
Realized returns are adjusted to account for transaction costs, with the total cost rate set at 0.3\%. Key performance metrics, including cumulative return, annualized return, Sharpe ratio, and maximum drawdown, are calculated over the full horizon of backtesting.

\begin{figure}[H]
    \centering
    \includegraphics[width=\textwidth]{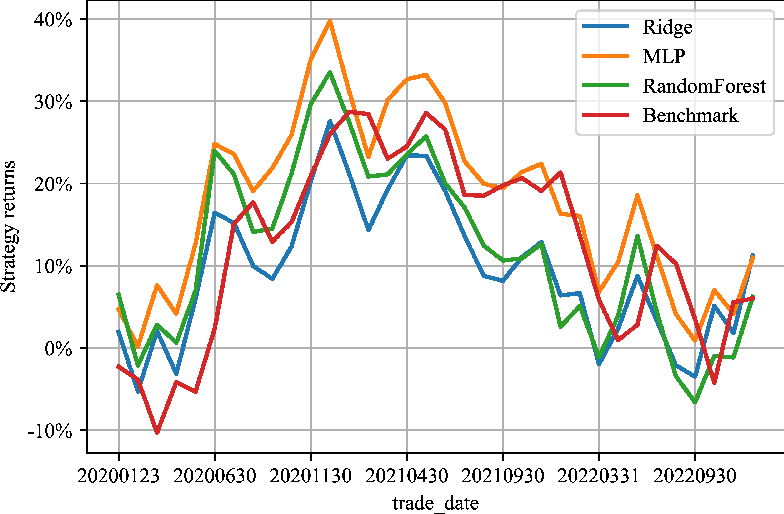}
    \caption{Strategy returns of three machine learning algorithms}
    \label{strategy returns 3}
\end{figure}

The backtesting period spans from January 1, 2020, to December 31, 2022. The CSI 300 Index is used as the benchmark for performance evaluation. The results are summarized in Table~\ref{backtest_results_3} and Figure~\ref{strategy returns 3}.

\begin{algorithm}[H]
\caption{Rolling Forecast for Stock Selection with Single Machine Learning}
\KwIn{
    $\mathcal{M} = \{M_{\text{Ridge}}, M_{\text{MLP}}, M_{\text{Random Forest}}\}$: Set of prediction models;\\
    $X_t$: Feature matrix at time $t$;\\
    $y_t$: Actual returns at time $t$;\\
    Training window size: 12 months;\\
    Test window size: 1 month;\\
    Backtest period: January 2020 to December 2022;\\
    Transaction cost rate: 0.3\%;
}
\KwOut{
    $\hat{y}_t^{a}$: Predicted returns for stock $a$ at time $t$;\\
    Backtesting performance metrics;
}

\For{$t = \text{start month} + 12$ months \KwTo end month}{
    Construct training set $\{(X_\tau, y_\tau)\}$ for $\tau = t-12$ to $t-1$\;
    Use $X_t$ as the test set\;
    \ForEach{model $M_i \in \mathcal{M}$}{
        Train $M_i$ using the training set\;
        Predict $\hat{y}_t^{a}$ using $M_i$ on $X_t$\;
    }
    Form portfolio based on predicted returns $\hat{y}_t^{a}$\;
    Adjust realized returns by applying transaction costs (0.3\%)\;
    Record realized returns and evaluation metrics for month $t$\;
}
Aggregate backtesting results over the entire period\;
\end{algorithm}

\FloatBarrier
\begin{table}[htbp]
    \setlength{\abovecaptionskip}{0cm}
    \caption{Backtest results of three machine learning algorithms}
    \label{backtest_results_3}
    \centering
    \setlength\tabcolsep{4pt} 
    \begin{tabularx}{\textwidth}{@{\extracolsep{\fill}} p{1.2cm} p{1.2cm} p{1.2cm} p{1.4cm} p{1.2cm} p{1cm} p{1cm} p{1cm} p{1.4cm}}
        \toprule
        & Strategy Return & Annualized Return & Annualized Volatility & Excess Return & Sharpe & Beta & Alpha & Maximum Drawdown \\
        \midrule
        Ridge          & 11.26\% & 3.98\% & 19.81\% & 5.23\% & 18.58\% & 113.56\% & 1.60\% & 48.69\% \\
        MLP            & 10.97\% & 3.88\% & 20.13\% & 4.94\% & 17.77\% & 117.34\% & 1.43\% & 53.74\% \\
        Random Forest  & 6.23\%  & 2.20\% & 21.66\% & 0.20\% & 8.78\%  & 135.81\% & -0.59\% & 74.34\% \\
        Benchmark      & 6.03\%  & 2.13\% & 18.59\% & 0.00\% & 9.86\%  & 100.00\% & 0.00\% & -- \\
        \bottomrule
    \end{tabularx}
\end{table}
\FloatBarrier

It is noteworthy that during the period from early 2020 to February 2021, the strategy returns generated by all three algorithms were generally higher than those of the benchmark. However, from March 2021 to September 2022, the returns of all three algorithms declined significantly and remained below the benchmark. These results suggest that the strategy based on a single machine learning algorithm may exhibit instability over time.

\begin{figure}[H]
    \centering
    \begin{minipage}{0.49\textwidth}
        \includegraphics[height=5.2cm,width=1\textwidth]{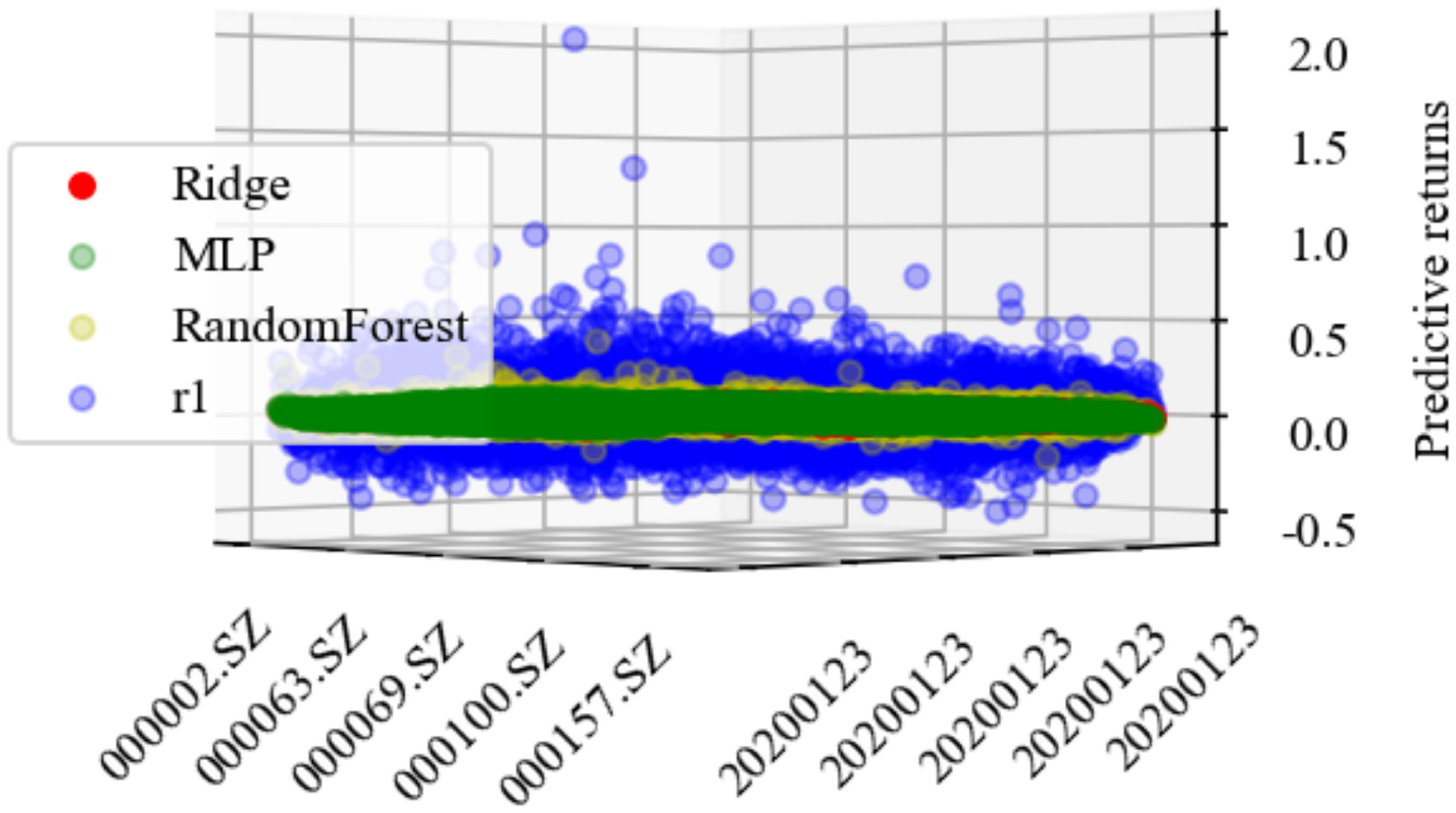}
        \caption{Predictive returns of three machine learning algorithms}
        \label{predictive returns}
    \end{minipage}
    \hfill  
    \begin{minipage}{0.48\textwidth}
        \includegraphics[width=1\textwidth]{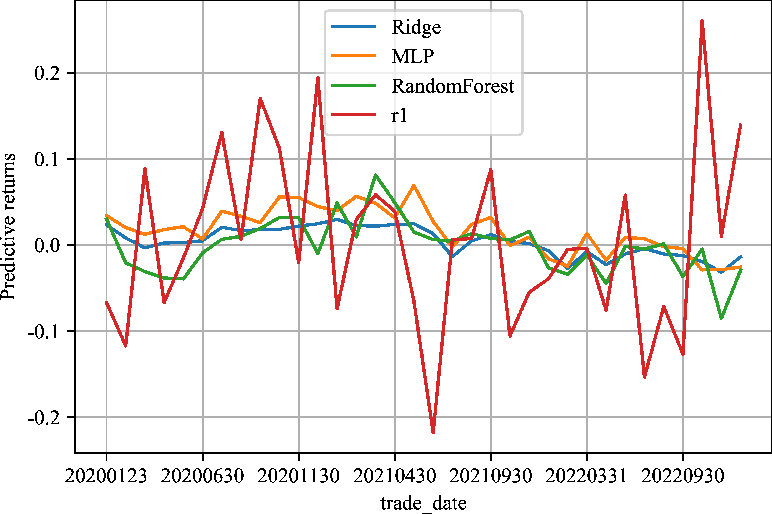}
        \caption{Predictive returns of three machine learning algorithms (000002.SZ as an example)}
        \label{predictive returns 2}
    \end{minipage}
\end{figure}

As shown in Figures~\ref{predictive returns} and~\ref{predictive returns 2}, the predictive return outputs of three machine learning algorithms are visualized, where $r_1$ denotes the actual next-period return. In Figure~\ref{predictive returns}, predictions by Ridge, MLP, and Random Forest are tightly clustered compared to the wider dispersion of actual returns, indicating generally conservative forecasts across stocks. In Figure~\ref{predictive returns 2}, which focuses on stock 000002.SZ, RidgeCV produces the most stable predictions, consistently smoothing over market fluctuations. Random Forest shows the largest variance, frequently overshooting or undershooting actual returns, while MLP offers a balance—tracking directional changes with moderate volatility. These results suggest that Ridge emphasizes stability, MLP captures trends reasonably well, and Random Forest may be more reactive in volatile market environments.

\begin{figure}[H]
    \centering
    \hspace*{\fill} 
    \begin{subfigure}[b]{0.3\textwidth}
        \centering
        \includegraphics[width=\textwidth]{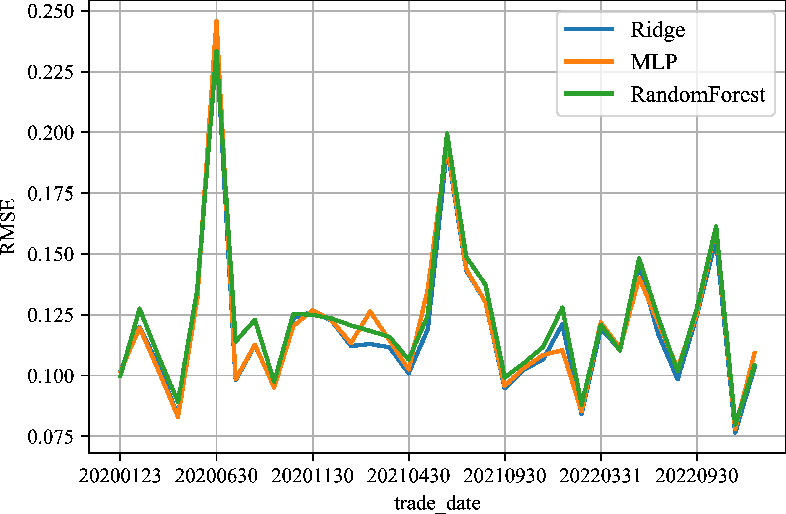}
        \caption{RMSE of the three algorithms}
    \end{subfigure}
    \hfill
    \begin{subfigure}[b]{0.3\textwidth}
        \centering
        \includegraphics[width=\textwidth]{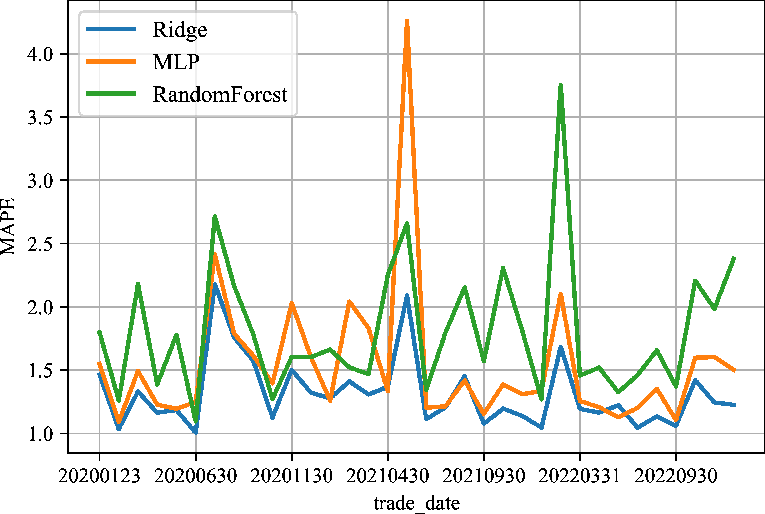}
        \caption{MAPE of the three algorithms}
    \end{subfigure}
    \hspace*{\fill}
    
    \vspace{1em}
    
    \begin{subfigure}[b]{0.3\textwidth}
        \centering
        \includegraphics[width=\textwidth]{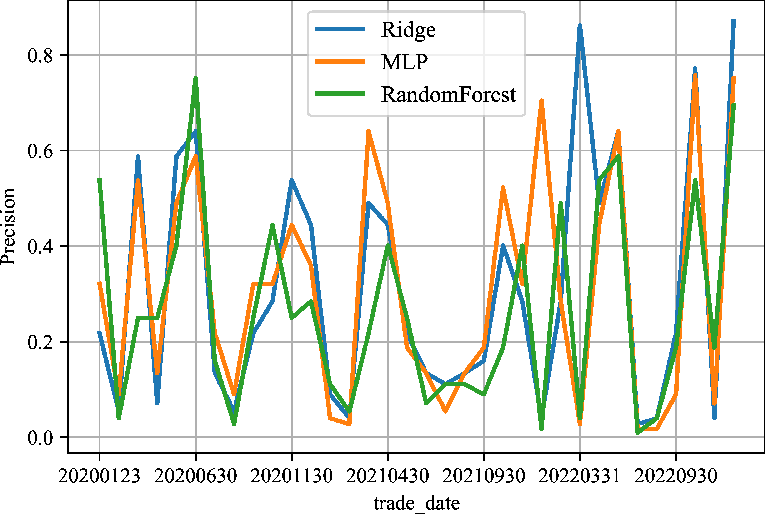}
        \caption{Precision of the three algorithms}
    \end{subfigure}
    \hfill
    \begin{subfigure}[b]{0.3\textwidth}
        \centering
        \includegraphics[width=\textwidth]{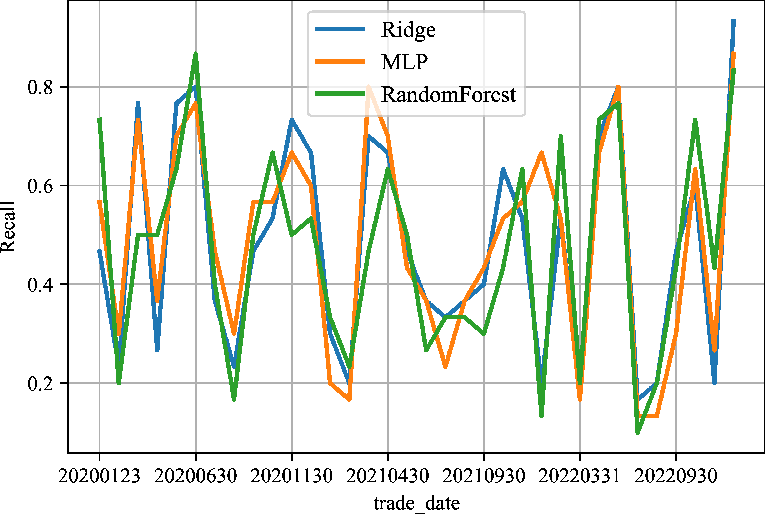}
        \caption{Recall of the three algorithms}
    \end{subfigure}
    \hfill
    \begin{subfigure}[b]{0.3\textwidth}
        \centering
        \includegraphics[width=\textwidth]{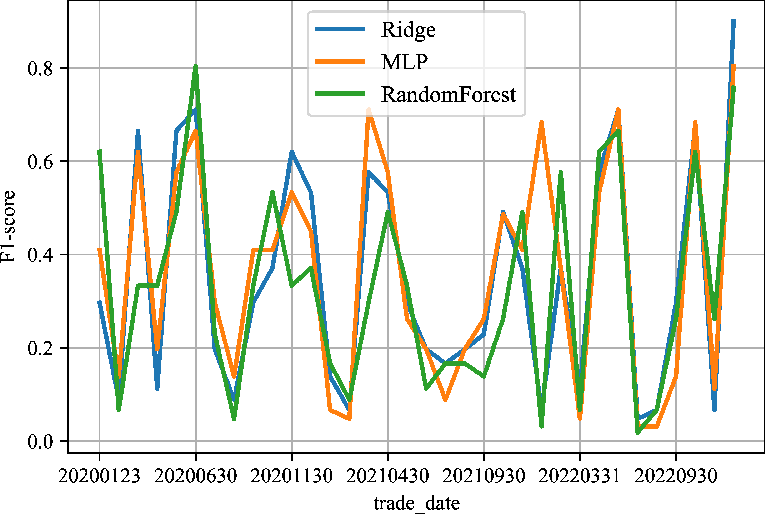}
        \caption{F1-score of the three algorithms}
    \end{subfigure}
    
    \caption{Five evaluation indicators of the three algorithms}
    \label{evaluation indicators 3}
\end{figure}

To compare the predictive performance of the three algorithms in the test data, two regression evaluation metrics-RMSE and MAPE, and three classification evaluation metrics—Precision, Recall and F1-score, are calculated. As shown in Figure~\ref{evaluation indicators 3}, all MAPE for three algorithms exceeds 1, indicating poor overall predictive performance. In addition,  the prediction accuracy of the three algorithms appears unstable, with an average accuracy of approximately 50\%. Further analysis of the regression metrics reveals that Ridge yields the lowest mean prediction error and the highest prediction accuracy. In contrast, Random Forest and MLP exhibit similar levels of prediction error. From the perspective of realized returns, Ridge demonstrates superior performance and is more likely to select stocks with higher actual returns, whereas Random Forest shows relatively weaker performance in this regard.

In next section, we propose a combined forecasting approach based on the three machine learning algorithms to enhance predictive performance and competitiveness.

\subsection{Forecasting with Combined Machine Learning Algorithms}

In this section, we conduct stock selection strategy based on combined machine learning algorithms. The key challenge is how to the determine the weights for the results by different algorithms. Here, we still apply the forehead three machine learning algorithms.Two types of weighting methods are introduced in the following text.

\subsubsection{Weighting based on Evaluation Metrics}

The weighting scheme based on evaluation metrics involves assigning weights to each algorithm according to regression or classification performance indicators. Regression metrics, such as RMSE and MAPE, evaluate the magnitude of prediction errors, where smaller values indicate better performance. Classification metrics, such as Precision, Recall and F1-score, measure the accuracy of predicting the direction of stock price movements, where higher values are preferred.

\begin{figure}[htbp]
    \centering
    \hspace*{\fill} 
    \begin{subfigure}[b]{0.3\textwidth}
        \centering
        \includegraphics[width=\textwidth]{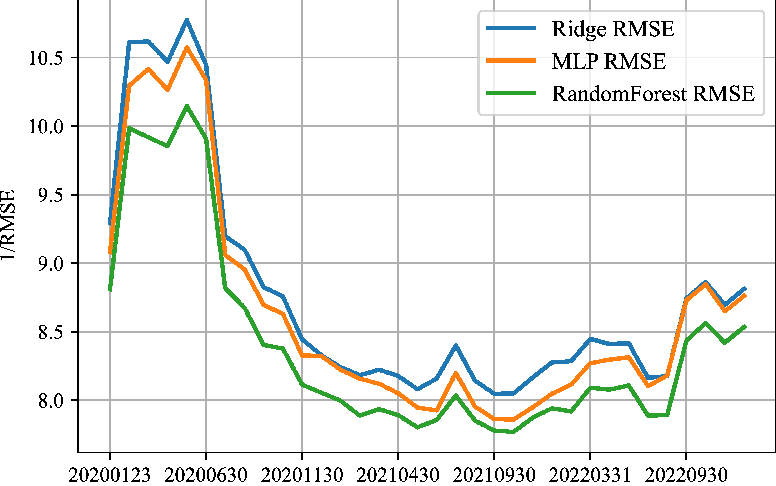}
        \caption{Weights based on 1/RMSE}
    \end{subfigure}
    \hfill
    \begin{subfigure}[b]{0.3\textwidth}
        \centering
        \includegraphics[width=\textwidth]{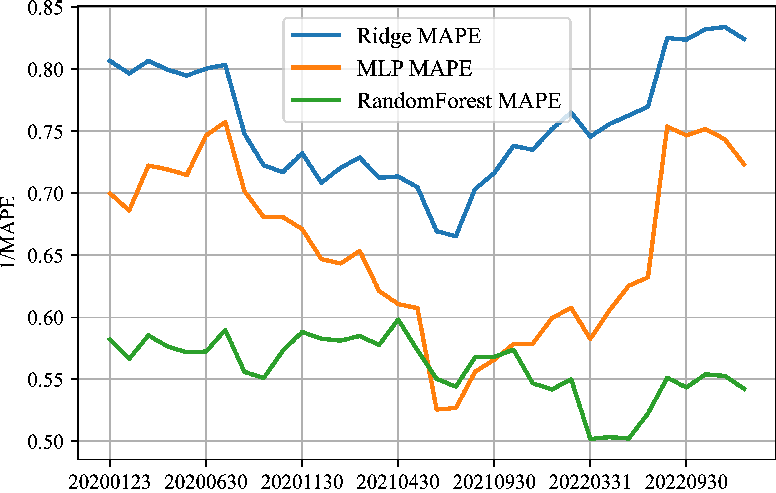}
        \caption{Weights based on 1/MAPE}
    \end{subfigure}
    \hspace*{\fill}
    
    \vspace{1em}
    
    \begin{subfigure}[b]{0.3\textwidth}
        \centering
        \includegraphics[width=\textwidth]{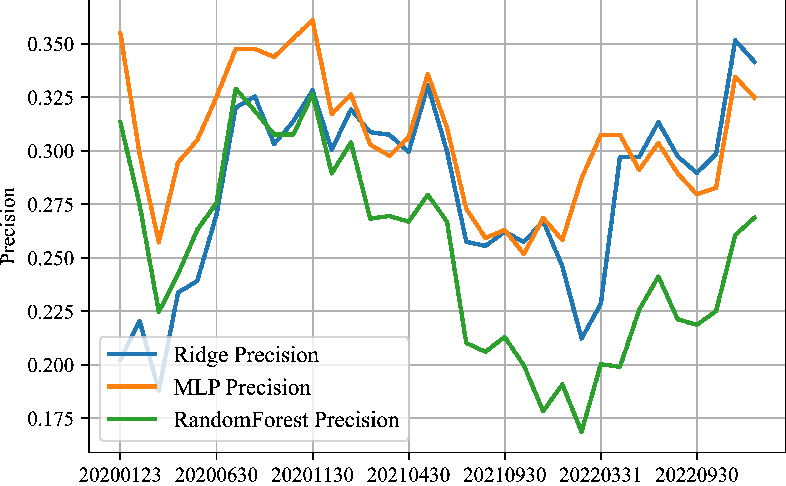}
        \caption{Weights based on precision }
    \end{subfigure}
    \hfill
    \begin{subfigure}[b]{0.3\textwidth}
        \centering
        \includegraphics[width=\textwidth]{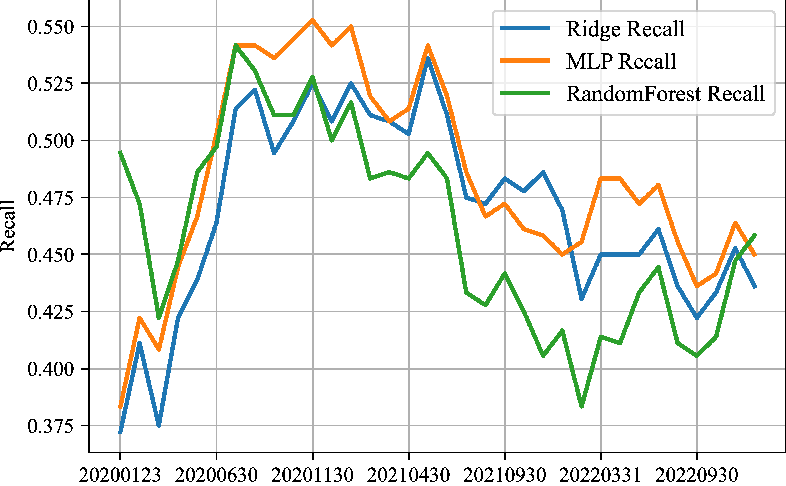}
        \caption{Weights based on recall}
    \end{subfigure}
    \hfill
    \begin{subfigure}[b]{0.3\textwidth}
        \centering
        \includegraphics[width=\textwidth]{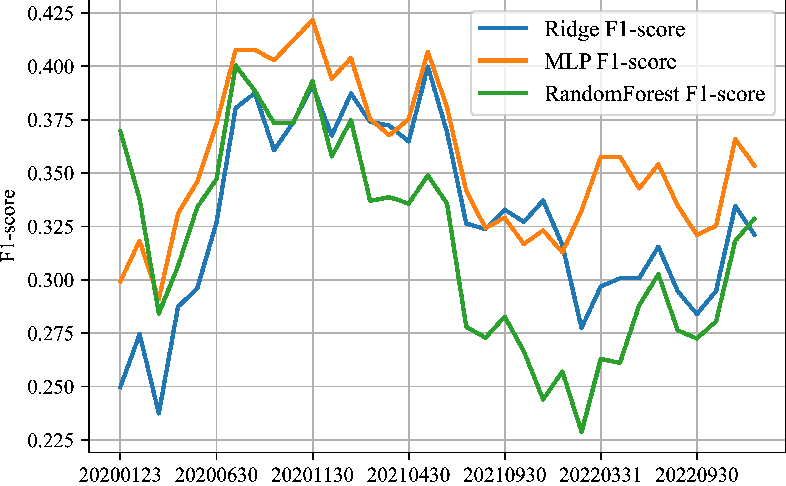}
        \caption{Weights based on F1-score}
    \end{subfigure}
    
    \caption{Combined prediction weights based on evaluation indicators (unstandardized)}
    \label{weights of evaluation indicators}
\end{figure}

To ensure the consistency in metric interpretation, the reciprocals of RMSE and MAPE are taken prior to weight calculation. The final weights are derived by normalizing the rolling average of each evaluation metric over the past 20 trading months. The unstandardized weights used in this method are illustrated in Figure~\ref{weights of evaluation indicators}.

As illustrated in Figure~\ref{weights of evaluation indicators}, the weights assigned to Random Forest and MLP based on RMSE and MAPE are similar and generally dominant, while Ridge receives a relatively lower weight. In terms of classification evaluation metrics, the weights assigned to the three algorithms vary over different time periods. From January 2020 to August 2021, MLP holds the highest weight, followed by Random Forest, with Ridge assigning the lowest. Since September 2021, the dominance gradually shifted toward Ridge and Random Forest.

\begin{figure}[htbp]
    \centering
    \includegraphics[width=\textwidth]{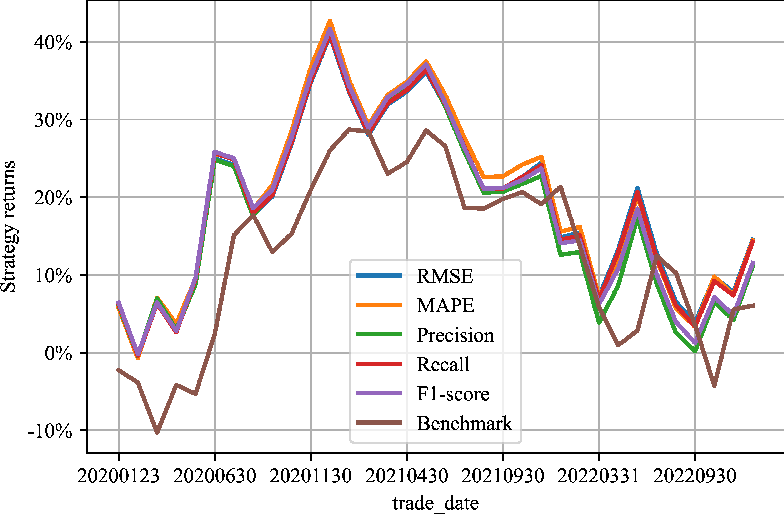}
    \caption{Strategy returns of combined forecasts based on evaluation metrics}
    \label{strategy returns 5}
\end{figure}

The backtesting results for the combined predictions based on the evaluation metrics are summarized in Table~\ref{backtest_results_5} and illustrated in Figure~\ref{strategy returns 5}.

\FloatBarrier
\begin{table}[htbp]
    \setlength{\abovecaptionskip}{0cm}
    \caption{Backtest results of combining forecasts based on evaluation metrics}
    \label{backtest_results_5}
    \centering
    \setlength\tabcolsep{4pt} 
    \begin{tabularx}{\textwidth}{@{\extracolsep{\fill}} p{1.2cm} p{1.2cm} p{1.2cm} p{1.4cm} p{1.2cm} p{1cm} p{1cm} p{1cm} p{1.4cm}}
        \toprule
        Weighting & Strategy Return & Annualized Return & Annualized Volatility & Excess Return & Sharpe & Beta & Alpha & Maximum Drawdown \\
        \midrule
        RMSE & 12.68\% & 4.48\% & 19.72\% & 6.65\% & 21.19\% & 112.62\% & 2.12\% & 3.77\% \\
        MAPE & 13.34\% & 4.71\% & 21.65\% & 7.31\% & 20.38\% & 135.73\% & 1.93\% & 3.98\% \\
        Precision & 16.50\% & 5.83\% & 20.62\% & 10.46\% & 26.80\% & 123.12\% & 3.27\% & 4.86\% \\
        Recall & 16.22\% & 5.73\% & 20.50\% & 10.19\% & 26.49\% & 121.65\% & 3.20\% & 4.86\% \\
        F1-score & 16.63\% & 5.88\% & 20.57\% & 10.60\% & 27.10\% & 122.52\% & 3.33\% & 5.44\% \\
        Benchmark & 6.03\% & 2.13\% & 18.59\% & 0.00\% & 9.86\% & 100.00\% & 0.00\% & 10.29\% \\
        \bottomrule
    \end{tabularx}
\end{table}
\FloatBarrier

As shown in Table~\ref{backtest_results_5}, the combined prediction weighted by F1-score achieves the best performance, with a strategy return of 16.63\%. In comparison, the benchmark return is only 6.03\%, and the corresponding maximum drawdown is limited to 5.44\%. Overall, combinations based on classification evaluation metrics consistently outperform those based on regression evaluation metrics.

It is worth noting that in the regression-based weighting schemes, Ridge consistently receives the lowest weight. However, under classification-based weighting, its weight becomes more comparable to those of the other two algorithms and even dominates during certain periods. This observation suggests that although Ridge may introduce a higher prediction bias in absolute value, it can still effectively identify stocks with significant upside potential under specific market conditions. When used in combination with other models under classification-based weighting, Ridge contributes positively to return performance through model synergy.

A comparison of the results in Table~\ref{backtest_results_3} and Table~\ref{backtest_results_5} reveals that the combined prediction based on evaluation metrics yielded backtesting results more than twice as favorable as those of any single algorithm.

\subsubsection{Weighting based on Information Coefficient }

Regression-based weighting primarily addresses numerical bias in return prediction, as captured by RMSE and MAPE. Such methods are commonly used in traditional financial forecasting but are less effective under volatile market conditions. In contrast, classification-based weighting methods focus on predicting the direction of price changes, rise or fall. These metrics are particularly suited for trend-following strategies and short-term market movements.

To address the limitations of regression- and classification-based weighting methods, a dynamic weighting method based on Information Coefficient (IC) is developed. The IC measures the cross-sectional correlation between predicted and actual returns, allowing both magnitude relevance and directional consistency to be captured. Higher IC values suggest stronger alignment between forecasts and market outcomes, and thus receive greater weights.

Originally introduced by Grinold (1989), IC is widely employed in multi-factor models to evaluate the effectiveness of predictive signals \cite{grinold1989fundamental}. However, its application to real-time dynamic weighting for combined machine learning algorithms, particularly for stock selection, is rarely investigated. In this framework, IC values are computed over a rolling window and used to assign time-varying model weights, without relying on additional forecasting models.

To implement this approach, two IC-based weighting schemes are adopted. The first is $IC_{Mean}$, which calculates the average IC value over a 20-trading-day rolling window and reflects the overall predictive ability of each model over recent periods. The second is the $IC_{Ratio}$, which adjusts the $IC_{Mean}$ by its standard deviation to account for the stability of predictive performance. 

For the $k$-th model, the IC at time $t$ is defined as
\begin{equation}
IC_{k}^{(t)} = \text{Corr}(\hat{R}_k^{(t)}, R^{(t+1)}),
\end{equation}
where $\hat{R}_k^{(t)}$ denotes the predicted return vector at time $t$ generated by model $k$ for all considered stocks, and $R^{(t+1)}$ is the actual realized return vector at the next time for all considered stocks. The correlation is calculated using Spearman’s rank correlation to reduce the influence of outliers.

\begin{itemize}
    \item \textbf{$IC_{Mean}$:}

    \begin{equation}
    IC_{{Mean}_k}^{(t)} = \frac{1}{L} \sum_{s=t-L}^{t-1} IC_{k}^{(s)},
    \end{equation}

    which summarizes the average predictive performance of model $k$ over a rolling window of length $L$. We take $L=20$ in this paper.

    \item \textbf{$IC_{Ratio}$:}

    \begin{equation}
    IC_{{Ratio}_k}^{(t)} = \frac{IC_{{Mean}_k}^{(t)}}{\sigma\left(IC_{k}^{(t-T):(t-1)}\right)},
    \end{equation}

    where $\sigma\left(IC_k^{(t-L):(t-1)}\right)$ is the standard deviation of IC values over the past $L$ periods, with $L= 20$.
\end{itemize}

\begin{algorithm}[H]
\caption{Dynamic Combined Forecasting Based on Information Coefficient (IC)}
\label{alg:ic_weighting}
\KwIn{
    \\
    \quad $\mathcal{M} = \{M_1, M_2, ..., M_k\}$: Set of prediction models;\\
    \quad $\hat{r}_{i,t}$: Predicted returns of model $M_i$ at time $t$;\\
    \quad $r_t$: Actual returns at time $t$;\\
    \quad $L$: Rolling window size (e.g., $L=20$);\\
    \quad Weighting scheme: IC-Mean or IC-IR
}
\KwOut{
    \\
    \quad $\hat{r}^{\text{combined}}_t$: Combined forecasted return at time $t$;\\
    \quad $w_{i,t}$: Weight assigned to model $M_i$ at time $t$
}

\For{$t = L+1$ \KwTo $T$}{
    \ForEach{model $M_i \in \mathcal{M}$}{
        Compute IC sequence in rolling window:\\
        \quad $IC_{i,\tau} = \text{SpearmanCorr}(\hat{r}_{i,\tau}, r_{\tau+1})$, for $\tau \in [t-L, t-1]$\\[0.5ex]
        
        Compute mean and std of IC:\\
        \quad $\mu_{IC_i} = \frac{1}{L} \sum_{\tau = t-L}^{t-1} IC_{i,\tau}$\\
        \quad $\sigma_{IC_i} = \text{Std}(IC_{i,\tau})$\\[0.5ex]
        
        Compute score based on chosen method:\\
        \quad $s_{i,t} =
        \begin{cases}
        \mu_{IC_i}, & \text{if IC-Mean weighting} \\
        \frac{\mu_{IC_i}}{\sigma_{IC_i} + \epsilon}, & \text{if IC-IR weighting}
        \end{cases}$
    }
    
    Normalize weights:\\
    \quad $S = \sum_{j=1}^{k} \max(s_{j,t}, 0)$\\
    \quad $w_{i,t} = \begin{cases}
    0, & \text{if } S = 0 \\
    \frac{\max(s_{i,t}, 0)}{S}, & \text{otherwise}
    \end{cases}$
    
    Compute combined forecast:\\
    \quad $\hat{r}^{\text{combined}}_t = \sum_{i=1}^{k} w_{i,t} \cdot \hat{r}_{i,t}$
}

\Return{$\{\hat{r}^{\text{combined}}_t\}_{t=N+1}^{T}$ and $\{w_{i,t}\}$}
\end{algorithm}

\begin{figure}[htbp]
    \centering
    \begin{subfigure}[b]{0.3\textwidth}
        \centering
        \includegraphics[width=\textwidth]{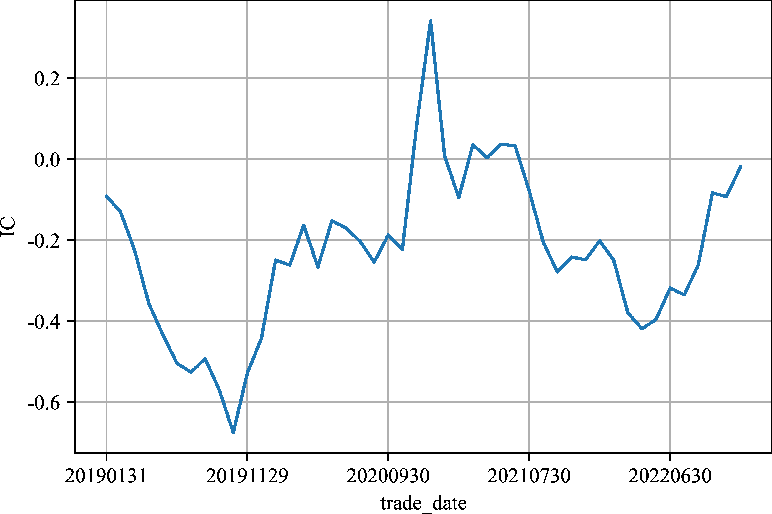}
        \caption{ Ridge }
    \end{subfigure}
    \hfill
    \begin{subfigure}[b]{0.3\textwidth}
        \centering
        \includegraphics[width=\textwidth]{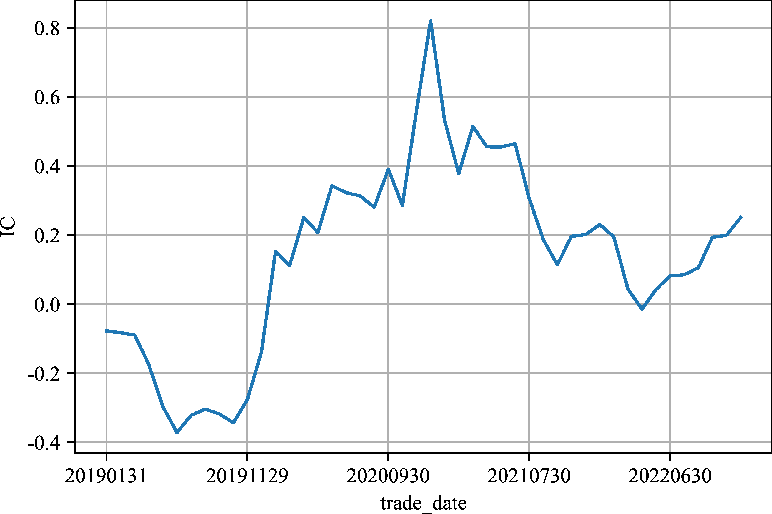}
        \caption{ MLP }
    \end{subfigure}
    \hfill
    \begin{subfigure}[b]{0.3\textwidth}
        \centering
        \includegraphics[width=\textwidth]{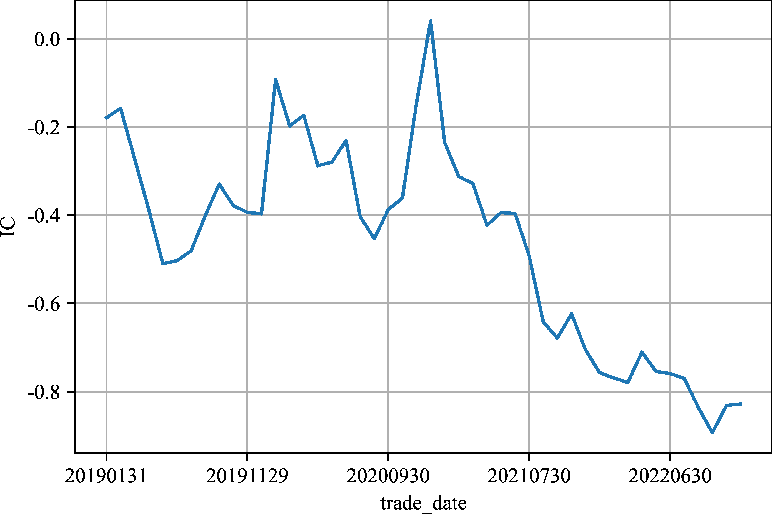}
        \caption{ Random Forest }
    \end{subfigure}
    \caption{Cumulative IC plots of the three machine learning predictors}
    \label{Cumulative IC}
\end{figure}

To monitor variations in predictive ability under different market environments, a dynamic time series of cumulative IC values is constructed. The trend of cumulative IC reflects the model's performance across distinct market phases and provides a basis for further optimization of dynamic weight allocation through the combined forecasting framework. The results are presented in Figure~\ref{Cumulative IC}. 

It is observed that the predictions generated by Random Forest exhibit a strong negative correlation with next-period returns. In contrast, MLP predictor demonstrates a moderately stronger positive correlation, while Ridge shows only a weak correlation overall.

\begin{figure}[H]
    \centering
    \begin{subfigure}[b]{0.48\textwidth}
        \centering
        \includegraphics[width=\textwidth]{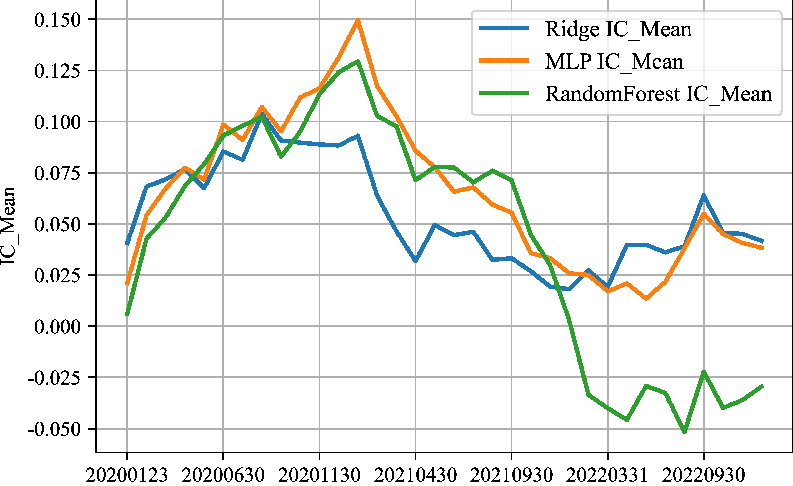}
        \caption{Weights Based on $IC_{Mean}$}
    \end{subfigure}
    \hfill
    \begin{subfigure}[b]{0.48\textwidth}
        \centering
        \includegraphics[width=\textwidth]{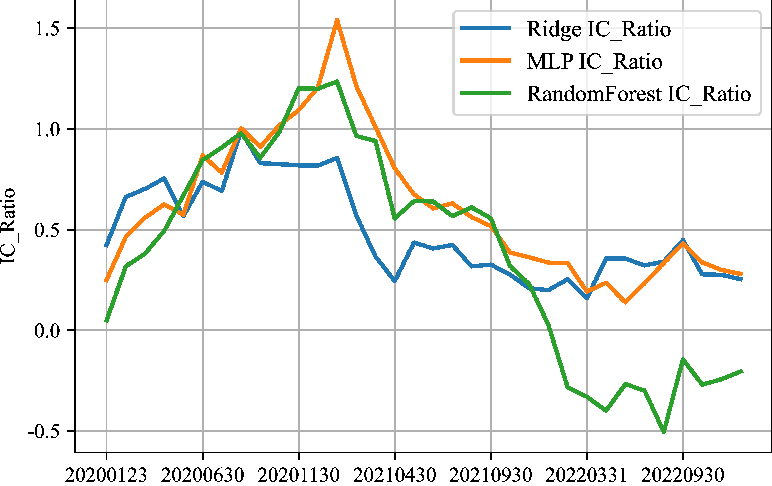}
        \caption{Weights Based on $IC_{Ratio}$}
    \end{subfigure}
    \caption{Unstandardized weights of the three machine learning algorithms based on IC}
    \label{weights of IC}
\end{figure}

The unstandardized IC-based weights assigned to the three machine learning algorithms are illustrated in Figure~\ref{weights of IC}.

As shown in Figure~\ref{weights of IC}, the weights derived from $IC_{Mean}$ indicate that the three algorithms receive similar weights prior to September 2020. Between September 2020 and January 2022, dominance alternates between Random Forest and MLP, while from February 2022 onward, Ridge and MLP alternate in dominance. The weights based on $IC_{Ratio}$ exhibit similar dynamics, with significant shifts observed in response to changes in the stability of the prediction models. These trends suggest that the IC-based weighting approach more effectively differentiates the predictive performance of the individual algorithms over time.

\begin{figure}[H]
    \centering
    \includegraphics[width=\textwidth]{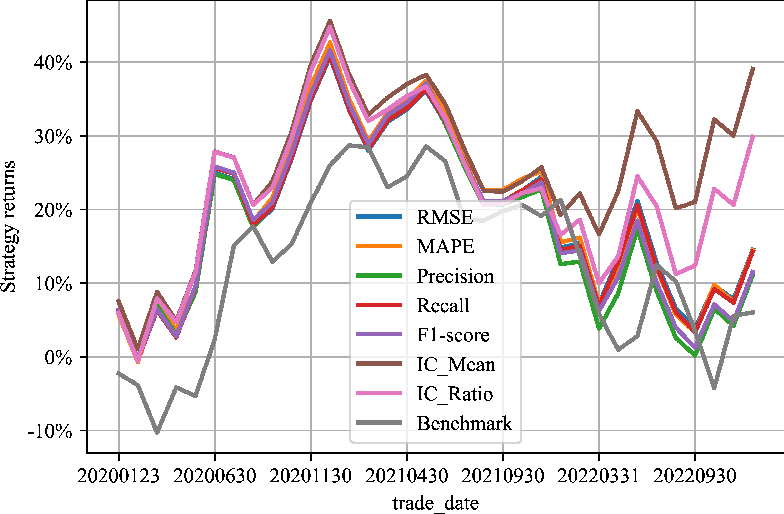}
    \caption{Strategy returns of combined prediction}
    \label{strategy returns 7}
\end{figure}

The backtesting results of combined prediction models with different weighting methods are summarized in Table~\ref{backtest_results_7} and Figure~\ref{strategy returns 7}.

\FloatBarrier
\begin{table}[htbp]
    \setlength{\abovecaptionskip}{0cm}
    \caption{Summary of backtesting results of combined forecast with different weighting methods}
    \label{backtest_results_7}
    \centering
    \setlength\tabcolsep{4pt} 
    \begin{tabularx}{\textwidth}{@{\extracolsep{\fill}} p{1.2cm} p{1.2cm} p{1.2cm} p{1.4cm} p{1.2cm} p{1cm} p{1cm} p{1cm} p{1.4cm}}
        \toprule
        Weighting & Strategy Return & Annualized Return & Annualized Volatility & Excess Return & Sharpe & Beta & Alpha & Maximum Drawdown \\
        \midrule
        RMSE       & 14.54\% & 5.14\% & 21.15\% & 8.51\% & 22.87\% & 129.49\% & 2.47\% & 67.47\% \\
        MAPE       & 14.45\% & 5.11\% & 21.41\% & 8.42\% & 22.44\% & 132.71\% & 2.37\% & 66.16\% \\
        Precision  & 11.11\% & 3.93\% & 21.60\% & 5.08\% & 16.78\% & 135.09\% & 1.15\% & 66.26\% \\
        Recall     & 14.27\% & 5.04\% & 21.31\% & 8.24\% & 22.25\% & 131.50\% & 2.33\% & 67.37\% \\
        F1-score   & 11.52\% & 4.07\% & 21.20\% & 5.49\% & 17.79\% & 130.16\% & 1.39\% & 67.37\% \\
        $IC_{Mean}$   & 39.09\% & 13.80\% & 22.07\% & 33.06\% & 61.17\% & 140.94\% & 10.92\% & 70.20\% \\
        $IC_{Ratio}$   & 29.89\% & 10.55\% & 22.13\% & 23.86\% & 46.33\% & 141.78\% & 7.66\% & 68.48\% \\
        Benchmark  & 6.03\%  & 2.13\%  & 18.59\% & 0.00\%  & 9.86\%  & 100.00\% & 0.00\% & -- \\
        \bottomrule
    \end{tabularx}
\end{table}
\FloatBarrier

It is evident that the backtesting results of the IC-based combined prediction approach significantly outperform those obtained through evaluation metrics-based weighting methods.
For example, the strategy returns based on IC\_{Mean} and IC\_{IR} stand at 39.09\% and 29.89\%, respectively. These results are more than double the returns generated by strategies based on regression evaluation metrics. Moreover, they are over four times the returns achieved by strategies relying on the Precision metric.

To assess and compare the predictive capabilities of diverse combination weighting methodologies, we employ a rolling window cross-validation approach. This involves utilizing a 12 month training period followed by a 1 month testing window, with the entire process being updated on a monthly basis. Such a setup enables us to effectively capture the dynamic and ever changing time series characteristics inherent in financial data. Our evaluation encompasses a range of weighting strategies, such as weighting based on RMSE, weighting derived from F1-score, and IC-based dynamic weighting.

\FloatBarrier
\begin{table}[htbp]
\centering
\caption{Rolling window cross-validation results across weighting methods}
\label{tab:rolling_window_comparison}
\begin{tabularx}{\textwidth}{l *{7}{>{\centering\arraybackslash}X}}
\toprule
\diagbox{Metrics}{Weighting} & RMSE & MAPE & Precision & Recall & F1-score & $IC_{Mean}$ & $IC_{Ratio}$ \\
\midrule
RMSE       & 0.119880 & 0.119765 & 0.119831 & 0.119887 & 0.119871 & 0.127404 & 0.132156 \\
MAPE       & 1.466802 & 1.449116 & 1.458110 & 1.467567 & 1.464411 & 1.852460 & 1.905561 \\
Precision  & 0.279435 & 0.280921 & 0.276405 & 0.278627 & 0.275289 & 0.321143 & 0.319450 \\
Recall     & 0.483333 & 0.479412 & 0.476471 & 0.481373 & 0.478431 & 0.508824 & 0.512745 \\
F1-score   & 0.344545 & 0.343850 & 0.339661 & 0.343274 & 0.339781 & 0.377248 & 0.379314 \\
IC         & 0.050745 & 0.052381 & 0.053182 & 0.050955 & 0.051960 & 0.056615 & 0.053669 \\
\bottomrule
\end{tabularx}
\end{table}
\FloatBarrier

As shown in Table~\ref{tab:rolling_window_comparison}, the IC-based models—particularly $IC_{Mean}$ and $IC_{Ratio}$—achieve the highest Recall (0.5088 and 0.5127), F1-score (0.3772 and 0.3793), and IC values (0.0566 and 0.0537), indicating stronger signal quality and predictive consistency. Although MAPE is somewhat higher (1.8525 and 1.9056), the overall results suggest improved risk-adjusted performance under varying market conditions.

\FloatBarrier
\begin{table}[htbp]
\centering
\caption{Results of $IC_{Mean}$ weighted predictor before and after screening factors}
\label{backtest_results_screening}
\setlength\tabcolsep{4pt} 
\begin{tabular*}{\textwidth}{@{\extracolsep{\fill}} p{1.2cm} p{1.2cm} p{1.2cm} p{1.4cm} p{1.2cm} p{1cm} p{1cm} p{1cm} p{1.4cm}}
\toprule
$IC_{Mean}$ weighted predictor & Strategy Return & Annualized Return & Annualized Volatility & Excess Return & Sharpe & Beta & Alpha & Maximum Drawdown \\
\midrule
After screening & 39.09\% & 13.80\% & 22.07\% & 33.06\% & 61.17\% & 140.94\% & 10.92\% & 70.20\% \\
Before screening & 16.65\% & 5.88\% & 22.99\% & 10.62\% & 24.28\% & 153.01\% & 2.78\% & 68.28\% \\
\bottomrule
\end{tabular*}
\end{table}
\FloatBarrier

In addition, we compare the backtesting results by  $IC_{Mean}$-based weighting method  before and after factor screening. See 
Table~\ref{backtest_results_screening}. A substantial improvement in strategy performance is observed following the elimination of low-quality factors. Specifically, the strategy return increases from 16.65\% to 39.09\%, and the Sharpe ratio rises from 24.28\% to 61.17\%, indicating enhanced risk-adjusted returns. While the annualized volatility remains relatively stable, alpha increases from 2.78\% to 12.96\%, suggesting an improved ability to generate excess returns independent of market movements. The maximum drawdown shows only a slight increase, implying that the higher returns are not driven by significantly higher risk. These results demonstrate that factor screening enhances the model’s capacity to identify stocks with greater return potential, thereby improving overall strategy performance.

\section{Conclusion and Discussion}
\label{sec4}
In this paper, we propose a stock selection strategy based on combined machine learning (ML) techniques with dynamic weighting methods. In terms of feature selection, we construct 17 novel second-level style factors derived from the CNE5 model to better capture emerging risk premia in financial markets and dynamic investor behaviors. These factors are designed to reflect market anomalies and behavioral biases that are not fully explained by traditional risk factors.

Three representative ML algorithms, Ridge Regression, MLP, and Random Forest, are employed to forecast stock returns. To mitigate the instability of single-algorithm prediction, we provide two types of combined prediction methods. (1) Evaluation metric-based weighting: Aggregates predictions using static weights derived from model performance metrics (e.g., RMSE, MAPE, Precision, Recall and F1 score );
(2)IC-based weighting: Implements dynamic weighting to reflect both the magnitude of prediction errors and the directionality of price movements. This approach adjusts weights iteratively based on the time-varying predictive power of each algorithm.
Empirical results demonstrate that the IC-based weighting method significantly outperforms the evaluation metric-based counterpart in terms of backtested returns, predictive accuracy, and risk-adjusted performance. Specifically, the $IC_{mean}$-weighted predictor achieves the highest competitiveness, yielding an annualized return of 13.80\% and an excess return of 39.09\% relative to the CSI 300 benchmark. Moreover, after refining the factor set by eliminating ineffective factors via Lasso regression (to address multicollinearity), the backtested performance of the $IC_{mean}$-weighted strategy improves further, indicating the importance of factor selection in model robustness. Notably, this dynamic weighting framework also holds great potential for improving predictive modeling and decision-making performance in other domains beyond finance, such as healthcare analytics, macroeconomic forecasting, and energy load prediction.

In the future, we plan to improve our quantitative stock selection strategies through two key approaches: adopting more advanced machine learning algorithms and enhancing model interpretability. To develop a more competitive hybrid algorithm, we could integrate deep learning methods (e.g., LSTM, TabNet, Transformers) to better capture temporal dependencies and complex interactions in financial data. Additionally, by incorporating regime-switching analysis, we can leverage SHAP (SHapley Additive exPlanations) and LIME (Local Interpretable Model-Agnostic Explanations) to interpret predictions and quantify feature importance for strategy returns.

\subsection*{Author Contributions}
Conceptualization, C.Z.; methodology, L.C.; software, L.C. and Z.H.; validation, L.C.; formal analysis, L.C.; investigation and resources, Z.H.; visualization, L.C. and Z.H.; writing—original draft preparation, L.C.; writing—review and editing, Z.H. and C.Z.; supervision, C.Z. All authors have read and agreed to the published version of the manuscript.

\subsection*{Data Availability}
 The data employed in this study are obtained from publicly accessible databases\cite{tushare2023data} \cite{swsresearch2023data}. You can ask for access to more specific data for verification reasons by asking the corresponding author.

\subsection*{Declarations}
  \textbf{Competing interests}: We declare that we have no known competing financial interests or personal relationships that could have appeared to influence the work reported in this paper.\\
   \textbf{Funding}: No financial support is received for this study.\\
  \textbf{Ethical approval}: This study does not involve human or animal subjects and therefore requires no ethical approval.

\bibliographystyle{sn-mathphys} 

\end{document}